\documentclass[prl,twocolumn,superscriptaddress,altaffillsymbol]{revtex4-1}
\usepackage{graphicx}
\usepackage{epstopdf}
\usepackage{color}
\usepackage{siunitx}
\usepackage{float}
\usepackage{amsmath}
\begin{document}

\title{Locating the missing superconducting electrons in the overdoped cuprates La$_{2-x}$Sr$_{x}$CuO$_{4}$}

\author{Fahad Mahmood}
\email{fahad@jhu.edu}
\affiliation{The Institute for Quantum Matter, Department of Physics and Astronomy, The Johns Hopkins University, Baltimore, MD 21218 USA.}
 
\author{Xi He}
\affiliation{Brookhaven National Laboratory, Upton, NY 11973, USA.}
\affiliation{Applied Physics Department, Yale University, New Haven, Connecticut 06520, USA.}

\author{Ivan Bo{\v{z}}ovi{\'{c}} }
\affiliation{Brookhaven National Laboratory, Upton, NY 11973, USA.}
\affiliation{Applied Physics Department, Yale University, New Haven, Connecticut 06520, USA.}

\author{N.~P.~Armitage}
\email{npa@pha.jhu.edu}
\affiliation{The Institute for Quantum Matter, Department of Physics and Astronomy, The Johns Hopkins University, Baltimore, MD 21218 USA.}


\begin{abstract}
Overdoped high temperature cuprate superconductors have often been understood within the standard BCS framework of superconductivity. However, measurements in a variety of overdoped cuprates, indicate that the superfluid density is much smaller than expected from BCS theory and decreases smoothly to zero as the doping is increased. Here, we combine time-domain THz spectroscopy with kHz range mutual inductance measurements on the \textit{same} overdoped La$_{2-x}$Sr$_{x}$CuO$_{4}$ films to determine the total, superfluid and uncondensed spectral weight as a function of doping. A significant fraction of the carriers remains uncondensed in a wide Drude-like peak as $T\rightarrow0$ while the superfluid density remains linear-in-temperature. These observations are seemingly inconsistent with existing, realistic theories of impurity scattering suppressing the superfluid density in a BCS-like \textit{d}-wave superconductor. Our large measurement frequency range gives us a unique look at the low frequency spectral weight distribution, which may suggest the presence of quantum phase fluctuations as the critical doping is approached.	
\end{abstract}

\maketitle
\setlength\belowcaptionskip{-3ex}

Unlike their underdoped counterparts \cite{Keimer_Nat_2015}, overdoped cuprate superconductors have been believed to be well described in terms of conventional BCS-like physics because of their relatively high carrier density \cite{Emery_Nat_1995} and the observation of a large and well defined Fermi surface in both photoemission (e.g., \cite{Plate_PRL_2005}) and quantum oscillation (e.g., \cite{Vignolle_Nat_2008}) experiments. While these observations imply that the normal state is conventional, it is an open question whether the superconducting state is conventional or not. Indeed, other studies indicate anomalies e.g., it has been found that the superfluid density was lower than expected as seen in different families of overdoped cuprates (e.g., Tl-2201 \cite{Uemura_Nat_1993,Niedermayer_PRL_1993}, Hg-1201\cite{Panagopoulos_PRB_2003}, Bi-2212 \cite{Corson_PhysicaB_2000}, La-214 \cite{Locquet_PRB_54,Lemberger_PRB_2010,Lemberger_PRB_2011,Rourke_Natphys_2011}), and most recently shown comprehensively in overdoped La$_{2-x}$Sr$_{x}$CuO$_{4}$ films \cite{Bozovic_Nat_2016}.

The unexpectedly low superfluid density naturally leads to two important questions: (1) Where are the ``missing'' carriers that do not condense into the superfluid? And (2) why do they not condense? These issues are at the heart of the superconductivity debate in overdoped cuprates. While some of the past observed behavior may indicate pair-breaking or disorder affects within a BCS-like theory, conflicting ideas include electronic phase separation or the presence of large superconducting phase fluctuations and so a complete consensus has yet to emerge.

To study these questions, we utilize time-domain THz spectroscopy (TDTS) in conjunction with kHz range mutual inductance measurements to systematically track both the condensate and the free carrier spectral weight as a function of doping for overdoped La$_{2-x}$Sr$_{x}$CuO$_{4}$ films. We find that a significant fraction of the total spectral weight remains uncondensed as $T\rightarrow0$ and manifests as a Drude-like peak at frequencies comparable to the theoretical weak coupling BCS gap. Taken with the linearity of the superfluid density with temperature, our observations are difficult to reconcile with extant theories of a BCS-type \textit{d}-wave superconductor in the presence of impurity scattering. Analysis of the frequency dependence of the spectral weight distribution points to the presence of significant quantum phase fluctuations. This limits any mean-field description of the superconducting transition for overdoped La$_{2-x}$Sr$_{x}$CuO$_{4}$.

Dynamical measurements such as TDTS independently determine both the real and imaginary parts of the frequency dependent conductivity $\sigma(\nu)$ at the relevant energy scales for superconductivity \cite{Corson_Nature_1999}. Measurements presented here were performed on 20 monolayers ($\sim\SI{13.2}{\nano\meter}$) thick La$_{2-x}$Sr$_{x}$CuO$_{4}$ films deposited on LaSrAlO$_4$ substrates by molecular-beam-epitaxy. Figure 1 shows the real $\sigma_1(\nu)$ and imaginary $\sigma_2(\nu)$ conductivities at different temperatures, $T$, for an overdoped La$_{2-x}$Sr$_{x}$CuO$_{4}$ film ($x = 0.23$) with $T_c =\SI{27.5}{\kelvin}$. For $T\gg T_c$, $\sigma_1(\nu)$ is frequency independent while $\sigma_2(\nu)$ is small, which is consistent with the behavior of a normal metal at frequencies well below the scattering rate. As the temperature is lowered across $T_c$, $\sigma_1(\nu)$ first rises and then decreases as spectral weight at higher frequencies is transferred to frequencies below the measurement range. Similarly, below $T_c$, $\sigma_2(\nu)$ develops a $1/\nu$-like dependence as the low frequency spectral weight condenses into a delta function at $\nu = 0$. However, even down to the lowest temperatures in the superconducting state ($T =\SI{1.6}{\kelvin}$), $\sigma_1(\nu)$ remains comparable in size to the normal state $\sigma_1(\nu)$ (Fig.\ 1b).  A similar residual $\sigma_1(\nu)$ as $T\rightarrow0$ has also previously been observed in films of the cuprate Bi-2212 for a range of dopings \cite{Corson_PRL_2000,Corson_PhysicaB_2000}. Such observations are incompatible with conventional BCS-like behavior in the absence of impurity scattering where nearly all the low frequency spectral weight should condense into a $\nu = 0$ delta function and consequently $\sigma_1(\nu)$ at THz frequencies should be negligible in the $T \rightarrow0$ limit.

\begin{figure}
	\includegraphics[width=1\columnwidth]{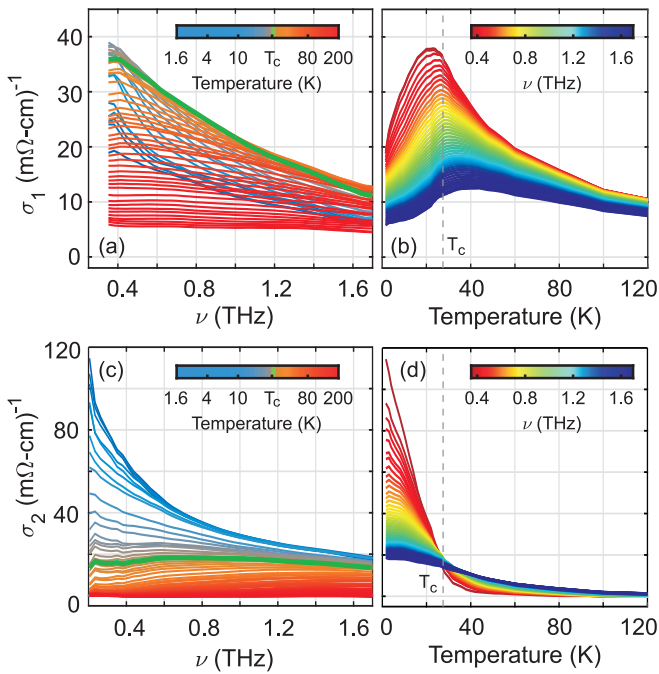}
	\caption{Real (\textbf{a},\textbf{b}) and imaginary (\textbf{c},\textbf{d}) parts of the THz optical conductivity as a function of frequency ($\nu$) and temperature. Green curves in (a) and (c) indicate the conductivities at $T_c$. Vertical dashed lines in (b) and (d) denote $T_c$.}
	\label{Fig1}
\end{figure}

\begin{figure*}
	\includegraphics[width=2.05\columnwidth]{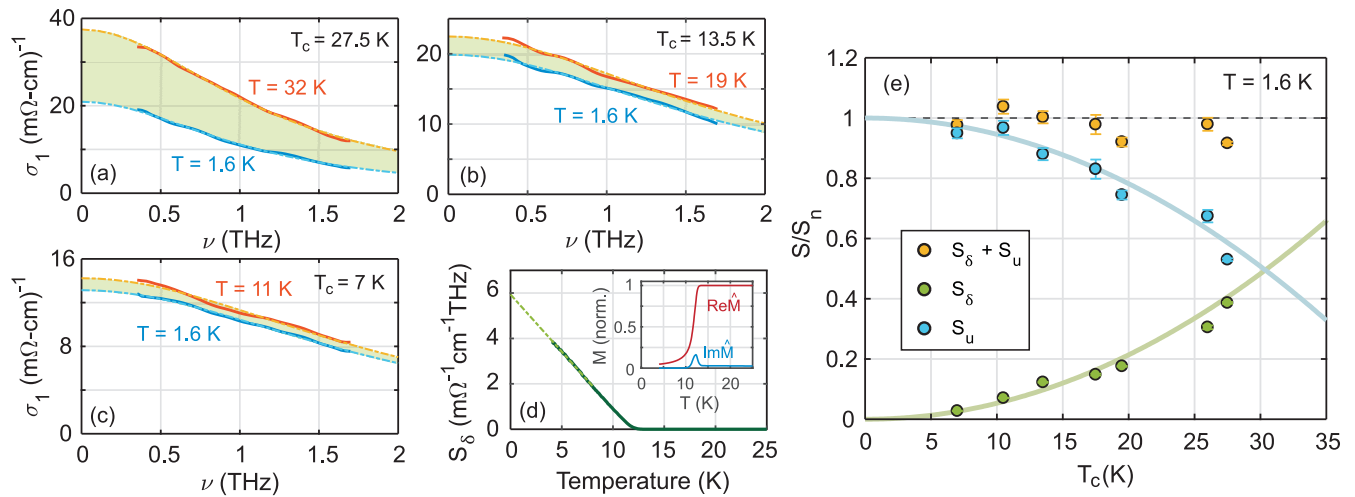}
	\caption{{$\sigma_1(\nu)$ above and below $T_c$ for overdoped La$_{2-x}$Sr$_x$CuO$_4$ films with \textbf{(a)} $T_c = \SI{27.5}{\kelvin} $, \textbf{(b)} $T_c = \SI{13.5}{\kelvin} $ and \textbf{(c)} $T_c = \SI{7}{\kelvin} $. Solid lines indicate the data. Dashed lines show a Drude fit with a single scattering rate, $\sigma_1(\nu) = S\tau/(1+\nu^2\tau^2)$. Shaded region represents the expected superfluid spectral weight. \textbf{(d)} The superfluid spectral weight $S_{\delta}$ with temperature for the film with $T_c = \SI{13.5}{\kelvin}$ as derived from the complex impedance using a two-coil MI setup ($\nu = \SI{40}{\kilo\hertz}$). The dashed line represents a linear extrapolation to determine $S_{\delta}$ for $T =\SI{1.6}{\kelvin}$ Inset: The real and imaginary parts of the mutual inductance with temperature for the same film. \textbf{(e)} Spectral weight, normalized to the normal state spectral weight $S_{n}$, of the superfluid ($S_{\delta}$) and of the uncondensed carriers ($S_{u}$) as a function of doping at $T = \SI{1.6}{\kelvin}$. $S_{\delta}$ is determined from the MI data as in (d) while $S_{u}$ and $S_{n}$ are determined from Drude fits to $\sigma_1(\nu)$. Yellow circles give $(S_{\delta} +S_{u})/ S_{n} $. Solid lines are guides to the eye. Error bars represent the $95\%$ confidence interval (2 s.d.) in the fitting procedure to extract $S$.}}
	\label{Fig1}
\end{figure*} 

To study this further, we directly compare the measured $\sigma_1(\nu)$ in the normal state and in the limit $T\rightarrow0$ for a range of overdoped La$_{2-x}$Sr$_{x}$CuO$_{4}$ films. Figures 2a, 2b and 2c show $\sigma_1(\nu)$ at temperatures above and below $T_c$ for the superconducting films with $T_c = \SI{27.5}{\kelvin}$, $\SI{13.5}{\kelvin}$ and $\SI{7}{\kelvin}$ respectively. The carrier spectral weight ($S$) contributing to the finite frequency conductivity is directly proportional to the area under the $\sigma_1(\nu) $ curve i.e., $\int_{0+}^{\infty} \sigma_1(\nu) d\nu=\frac{\pi}{2}S$. Based on this, it is apparent from Fig.\ 2a-c that a significant fraction of the normal state spectral weight ($S_{n}$) remains uncondensed at THz frequencies as $T\rightarrow0$. Moreover, the ratio of the uncondensed spectral weight ($S_{u}$) to the normal spectral weight becomes even greater (i.e., more anomalous) for the more overdoped films. This behavior can be quantified by fitting $\sigma_1(\nu)$ at each doping and temperature to a single Drude peak i.e., $\sigma_1(\nu) = S\tau/(1+\nu^2\tau^2)$) (dashed lines on Fig.\ 2a-c). Figure 2e shows that the ratio of the uncondensed spectral weight to the normal state spectral weight monotonously approaches unity as the critical doping is approached i.e., $S_{u}/S_{n} \rightarrow 1$ as $T_c \rightarrow0$.

This observation naturally answers the first question raised above i.e., where are the ``missing'' carriers? Our results here indicate that they remain uncondensed in a THz wide Drude-like peak down to $T = 0$. To corroborate this, we have performed two-coil mutual inductance (MI) measurements on the \textit{same} films to extract the spectral weight in the superconducting delta function ($S_{\delta}$). Figure 2d shows $S_{\delta}$ as a function of temperature as obtained from the penetration depth ($\lambda$) from MI measurements ($S_{\delta} =  \frac{1}{2\pi\mu_0\lambda^2}$) for the $x = 0.23$ film (see Supplementary Material (SM) for details). As expected from previous MI measurements on overdoped La$_{2-x}$Sr$_{x}$CuO$_{4}$ films \cite{Bozovic_Nat_2016}, $S_{\delta}(T)$ is essentially linear with $T$ down to the lowest temperature. We extrapolate the data to obtain $S_{\delta} $ at $T = \SI{1.6}{\kelvin}$ to directly compare with $S_{n} $ and $S_{u} $ obtained from TDTS at the same temperature for a range of dopings (Fig.\ 2e). In the context of the Ferrel-Glover-Tinkham (FGT) sum rule, $S_{n} = S_{\delta}  +S_{u}$. Consequently, if our measured $S_{u}$ is indeed due to the ``missing'' carriers, then $[S_{\delta}+S_{u}]/S_{n} = 1 $ regardless of doping. As shown in Fig.\ 2e $S_{\delta}/S_{n} \rightarrow0$ as $T_c \rightarrow0$ while $[S_{\delta}+S_{u}]/S_{n}  \approx 1$ (within $\pm10\%$) for all samples. Although a small amount of spectral weight may be transferred to high frequencies below $T_c $\cite{molegraaf2002superconductivity}, our analysis shows that the vast majority remains at low frequencies. Note that because $T_c$ is decreasing across the series of samples, the reduced temperature at $T = \SI{1.6}{\kelvin}$ increases. This thermal effect on $S_{\delta}$ is negligible when considering the overall decrease in $S_{\delta}$ with doping (\cite{Bozovic_Nat_2016}). 

Having located the ``missing" spectral weight, we consider a few reasons these charge carriers don't condense. One obvious possibility is pair-breaking scattering due to impurities which smears out the \textit{d}-wave node leading to nodal Bogoliubov quasi-particles and a suppression of the superfluid density $n_s$. Such pair-breaking in both the unitary and Born scattering limits within BCS theory have been studied extensively \cite{Hirschfeld_PRL_1993,Hirschfeld_PRB_1993,Hosseini_PRB_1999,Durst_PRB_2000,Broun_PRL_2007}, with the latter considered recently \cite{Lee-Hone_PRB_2017} to explain the suppression in superfluid density observed in \cite{Bozovic_Nat_2016}. Our results presented here seem to be inconsistent with these models as previously implemented for the following reasons. 

First, aside from the delta function at $\nu = 0$, $\sigma_1(\nu)$ for a dirty \textit{d}-wave superconductor may be composed of both a narrow low frequency Drude-like peak and -- if the normal state scattering rate is larger than the superconducting gap $2\Delta/h$ -- a part that is an increasing function of $\nu$ (e.g., \cite{Maki_PRB_1994,Tajima_PRB_2005}). For weak-coupling \textit{d}-wave BCS, $2\Delta = 4.28k_BT_c$ and thus, $2\Delta/h$ is expected to range from $\SI{0.62}{\tera\hertz}$ to $\SI{2.45}{\tera\hertz}$ for the films studied here i.e., mostly within the spectral range of our spectrometer. Yet, we do not observe any signatures of $2\Delta$ compatible with this theory in the measured $\sigma_1(\nu)$ (Fig.~2a-c and SM Fig.~S5). This implies that the superconducting gap may be larger than expected from weak-coupling BCS and remains reasonably large as the critical doping is approached, or the gap's signature is otherwise suppressed in the spectra. Additionally, while the $\nu\rightarrow0$ limit of the residual $\sigma_1(\nu)$ can be sizable within dirty \textit{d}-wave theory \cite{Durst_PRB_2000}, the corresponding frequency dependence of $\sigma_1(\nu)$, within existing theory, is not expected to be the simple form of the single Lorentzian that we observe (except in the unitary limit).

Second, it is expected that impurity scattering drives a change from the expected linear-$T$ behavior of $n_s$ for a clean \textit{d}-wave superconductor to a quadratic dependence at a crossover temperature $T^{**}$ for both unitary and Born scatterers \cite{Hirschfeld_PRL_1993,Hirschfeld_PRB_1993,Hosseini_PRB_1999,Broun_PRL_2007}.  Irrespective of the kind of scattering, within extant theory, $T^{**} $ reflects a frequency scale $\gamma$ that is roughly the width of the residual Drude peak $1/\tau$ in the limit $T \rightarrow 0$ as $T^{**} \simeq \gamma =  1/\tau$ \cite{Hirschfeld_PRB_1993}. Figure 3a shows the extracted scattering rate $\gamma =  1/\tau$ with temperature for all the films studied in this work. For each film, $\gamma \gtrsim T_c$ and thus $T^{**} $ should be $\gtrsim T_c$. This implies that $n_s$ should scale quadratically with $T$ for all $T < T_c$. On the contrary, in the present films as well as in the previous work on similarly grown films \cite{Bozovic_Nat_2016}, $n_s$ remains quite linear down to the lowest temperatures (Fig.\ 2d) and no crossover behavior is observed. While a recent work \cite{Lee-Hone_arxiv_2018} claims that one can reconcile the observed broad residual Drude with the linear-$T$ behavior of $n_s$, the underlying calculations incorporate an unphysical infinite number of infinitely weak scatterers (Born-limit) and assume a much higher $T_c$ for overdoped La$_{2-x}$Sr$_{x}$CuO$_{4}$ than what has been observed in both single-crystals and thin films.

 	\begin{figure*}[!ht]
 		\includegraphics[width=2.05\columnwidth]{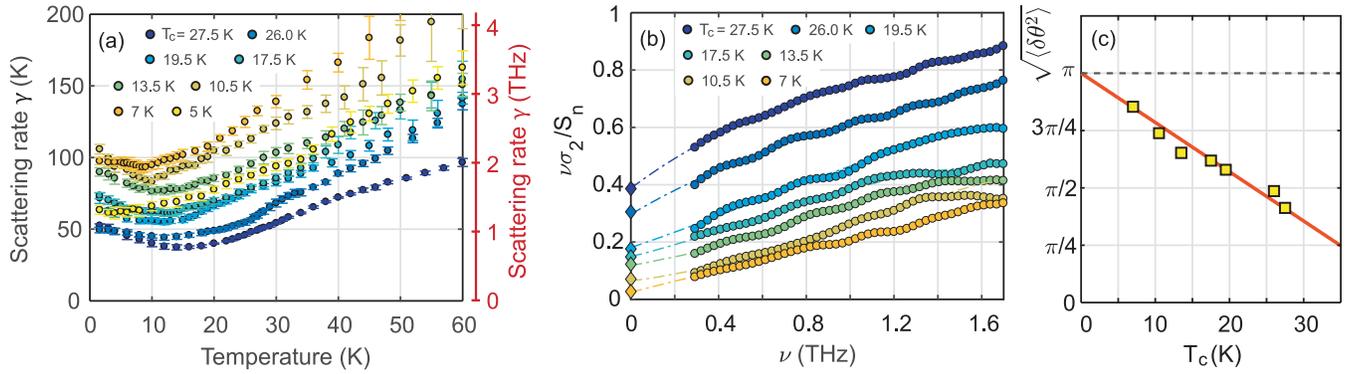}
 		\caption{\textbf{a} Scattering rate $\gamma$, in units Kelvin, with temperature for all films measured. $\gamma$ is obtained with a single Drude-fit to $\sigma_1(\nu)$ at all temperatures as shown in Fig.2 a-c. Error bars represent the $95\%$ confidence interval in the fitting procedure to extract the parameter $\gamma = 1/\tau$. \textbf{(b)} $\nu\sigma_2$, normalized to $S_{n}$, versus frequency for all films at $T = \SI{1.6}{\kelvin}$. Circle and diamond symbols represent the TDTS and MI data respectively. Dashed lines are guides to the eye. \textbf{(c)} The average phase uncertainty determined from the quantum Debye Waller factor $W_Q =  e^{-<\delta\theta^2>/2}$ as a function of doping. Here, $W_Q = S_{\delta}/S_{n}$. Red line is a linear guide to the eye to the data shown as yellow squares.}
 		\label{Fig1}
 	\end{figure*}

Observations reminiscent to ours have been made in heat capacity measurements of overdoped La$_{2-x}$Sr$_{x}$CuO$_{4}$ single crystals \cite{Wang_PRB_76}, where a large fermion-like linear-in-$T$ contribution was found deep into the superconducting state. For overdoped samples with $T_c \sim \SI{20}{\kelvin}$, the heat capacity coefficient was roughly $\SI{70}{\percent}$ of the normal state and reached essentially $\SI{100}{\percent}$ by $T_c \sim \SI{7}{\kelvin}$. A straightforward interpretation of such data is microscopic electronic phase separation i.e., the presence of superconducting regions embedded in a normal state metallic background. Our measurements support such a scenario in that the scattering rate of our $T\rightarrow0$ residual Drude is about the same as that in the normal state (Fig.~3a). However, if the residual Drude peak were due to such phase separation then the volume fraction corresponding to the normal metallic region needs to be exceedingly large (e.g., nearly $\SI{95}{\percent}$ for the film with $T_c = \SI{7}{\kelvin}$). It is hard to reconcile robust superconductivity as well as the exceedingly uniform $T_c$ for all the films measured (as characterized by a sharp transition in the dissipative part of the MI (Fig.~2d and \cite{Bozovic_Nat_2016})) with a scenario of such extreme phase separation.

A quantity that has been quite useful in understanding TDTS on cuprates is the phase stiffness $\mathcal{T}_\phi$, which is the energy scale to introduce twists in the phase $\phi$ of the superconducting order parameter $\Delta \mathrm{e}^{i\phi}$. As detailed in previous works \cite{Corson_Nature_1999,Bilbro_Natphys_2011,Bilbro11a}, $\mathcal{T}_\phi \propto \nu\sigma_2$. Measuring $\mathcal{T}_\phi$ at a finite frequency sets a length/time scale over which the system is dynamically probed. In the absence of fluctuations, the system will be stiff on all length and time scales and thus $\mathcal{T}_\phi\propto \nu\sigma_2$ should be independent of the probing frequency. In general, $\displaystyle{\lim_{\nu \to \infty}} \nu\sigma_2 = S_{n}$ i.e., the total spectral weight. Interpreting the residual Drude peak as uncondensed superconducting charge carriers allows us to study $\mathcal{T}_\phi$ as a function of frequency. We note that this is similar to the analysis performed in \cite{Corson_PRL_2000} where a residual conductivity peak in Bi-2212 was interpreted as arising from a superconducting collective mode. However, even in the absence of such an interpretation, analyzing  $\nu\sigma_2$ allows us to get a good picture of the distribution of the low frequency spectral weight.

Figure 3b shows $\nu\sigma_2$ (normalized to the normal state spectral weight $S_{n}$) at $T = \SI{1.6}{\kelvin}$ as a function of frequency for all films measured by TDTS. Note that the $\sigma_2$ considered here has been corrected from its measured value to take into account a small contribution from dielectric screening (SM). We also plot in Fig.\ 3b the relative superfluid spectral weight ($S_{\delta}/S_{n}$) as obtained from MI measurements at $\nu = \SI{40}{\kilo\hertz}$ on the same films to directly compare in the $\nu\rightarrow0$ limit. For all dopings, $\nu\sigma_2$ as measured in the THz region can be smoothly connected with the MI measurement. Moreover, $\nu\sigma_2$ is strongly increasing with probing frequency for all dopings. In systems where the entire $\sigma_2$ arises from superconducting correlations, such dependence indicates that the phase of the system appears ``stiffer'' when probed at higher frequencies (i.e., at shorter length and time scales) but fluctuations degrade the superconductivity on longer length and time scales. This perspective has been used previously to analyze the THz response of the thermally fluctuating regime above $T_c$ \cite{Corson_Nature_1999,Bilbro_Natphys_2011,Bilbro11a}. Additionally, the behavior of $\nu\sigma_2$ is in accordance with Kramers-Kronig relations for $\sigma$ (SM sec.~D).

Note that one naturally expects large quantum phase fluctuations to accompany the small $T\rightarrow0$ superfluid stiffness \cite{Bozovic_Nat_2016}. These phase fluctuations can either be seen as a consequence of the reduced superfluid density or as a cause. Distinguishing between the two is currently not possible within our analysis. Nevertheless, our measurement of the inductive response over a large frequency range allows us to perform a unique analysis to highlight the presence of quantum phase fluctuations. This analysis may hold independent of the mechanism for the suppression of the superfluid density. Renormalization of the system's diamagnetic response can be described in terms of a quantum Debye-Waller factor $W_Q$ \cite{Ioffe_Science_1999,Benfatto_PRB_2001,Kwon_PRL_2001} which has been used in theoretical works to parametrize the suppression of the superfluid density e.g., $W_Q= S_{\delta}/S_{n}$. In the context of the self-consistent harmonic approximation, $W_Q$ can be directly related to the root mean phase uncertainty of the order parameter as $W_Q = e^{-<\delta\theta^2>/2}$ \cite{Chakravarty_PRL_1986,Kivelson_PRL_1995}. This approach has been invoked to describe global phase coherence in Josephson-junction arrays \cite{Chakravarty_PRL_1986} and phase disordered \textit{s}-wave superconductors \cite{Mondal_PRL_2011}. This is an intermediate regime form that obviously cannot be valid in the critical regime itself. For the purpose of this analysis, we assume that the normal state $\sigma(\nu)$ just above $T_c$ gives the spectral weight of the normal state diamagnetic response ($S_{n}$). Figure 3c shows the root mean phase uncertainty $\sqrt{<\delta\theta^2>}$ determined from $W_Q$ ($S_{\delta}/S_{n}$ in Fig.\ 2e) as a function of doping. Remarkably, as the critical doping is approached ($T_c\rightarrow 0$), the average phase disorder $\sqrt{<\delta\theta^2>}$ extrapolates to $\pi$. The maximum $T_c$ for the La$_{2-x}$Sr$_{x}$CuO$_{4}$ cuprate family ($\sim\SI{45}{\kelvin}$) is near $\sqrt{<\delta\theta^2>}  \rightarrow 0$. Obviously $\pi$ is a significant natural scale for phase disordering at the transition and we propose that this is a key indicator of strong quantum phase fluctuations at the termination of the superconducting dome. 

We note that $T_c \propto \sqrt{S_{\delta}}$ is expected to emerge from quantum phase fluctuations within the (3+1)D-XY universality class with $z=1$. Indeed, this is the scaling found in \cite{Bozovic_Nat_2016}, but only very near the QCP, around $0.25 < x < 0.26$ (as well as in earlier work \cite{Lemberger_PRB_2011}). The same scaling was found for extreme underdoped YBCO, but only for $0.054 < x < 0.057 $ \cite{Broun_PRL_2007}. Note that all the data analyzed here are either extrapolated to $T=0$ or taken in a regime where they are temperature independent. Therefore, unlike previous work that concentrated on thermal superconducting fluctuations \cite{Corson_Nature_1999,Bilbro_Natphys_2011,Bilbro11a}, any fluctuations here are quantum in nature and are associated with the zero-point motion of the condensate. While our analysis points to the presence of quantum phase fluctuations, further experiments are needed to determine whether they cause the low superfluid density or not throughout the overdoped regime. Of course, close to the critical point fluctuations will indeed reduce the superfluid density and so this may be the regime where the critical scaling is observed. A scenario of phase fluctuations does not obviously explain features like the large linear-in-temperature heat capacity for overdoped samples \cite{Wang_PRB_76}. It could be that the ultimate picture needs to combine aspects of both fluctuations and phase separation where the transition proceeds through the phase disordering of weak superconducting links.

\vspace{0.2cm}

\begin{acknowledgments}
	The authors thank L.~Benfatto, D.~Broun, D.~Chaudhuri, D.~Chowdhury, S.~Dodge, P.~Hirschfeld, S.~Kivelson, M.~Mondal, C.~Varma, and  I.~Vishik, for helpful discussions. Research at JHU was funded by the US DOE, Basic Energy Sciences, Materials Sciences and Engineering Division through Grant DE-FG02-08ER46544. Film synthesis by MBE and characterization at BNL was supported by the US DOE, Basic Energy Sciences, Materials Sciences and Engineering Division. X.H. is supported by the Gordon and Betty Moore Foundation's EPiQS Initiative through Grant GBMF4410 to I.B.
\end{acknowledgments}

\bibliography{main} 

\begin{thebibliography}{34}%
\makeatletter
\providecommand \@ifxundefined [1]{%
 \@ifx{#1\undefined}
}%
\providecommand \@ifnum [1]{%
 \ifnum #1\expandafter \@firstoftwo
 \else \expandafter \@secondoftwo
 \fi
}%
\providecommand \@ifx [1]{%
 \ifx #1\expandafter \@firstoftwo
 \else \expandafter \@secondoftwo
 \fi
}%
\providecommand \natexlab [1]{#1}%
\providecommand \enquote  [1]{``#1''}%
\providecommand \bibnamefont  [1]{#1}%
\providecommand \bibfnamefont [1]{#1}%
\providecommand \citenamefont [1]{#1}%
\providecommand \href@noop [0]{\@secondoftwo}%
\providecommand \href [0]{\begingroup \@sanitize@url \@href}%
\providecommand \@href[1]{\@@startlink{#1}\@@href}%
\providecommand \@@href[1]{\endgroup#1\@@endlink}%
\providecommand \@sanitize@url [0]{\catcode `\\12\catcode `\$12\catcode
  `\&12\catcode `\#12\catcode `\^12\catcode `\_12\catcode `\%12\relax}%
\providecommand \@@startlink[1]{}%
\providecommand \@@endlink[0]{}%
\providecommand \url  [0]{\begingroup\@sanitize@url \@url }%
\providecommand \@url [1]{\endgroup\@href {#1}{\urlprefix }}%
\providecommand \urlprefix  [0]{URL }%
\providecommand \Eprint [0]{\href }%
\providecommand \doibase [0]{http://dx.doi.org/}%
\providecommand \selectlanguage [0]{\@gobble}%
\providecommand \bibinfo  [0]{\@secondoftwo}%
\providecommand \bibfield  [0]{\@secondoftwo}%
\providecommand \translation [1]{[#1]}%
\providecommand \BibitemOpen [0]{}%
\providecommand \bibitemStop [0]{}%
\providecommand \bibitemNoStop [0]{.\EOS\space}%
\providecommand \EOS [0]{\spacefactor3000\relax}%
\providecommand \BibitemShut  [1]{\csname bibitem#1\endcsname}%
\let\auto@bib@innerbib\@empty
\bibitem [{\citenamefont {Keimer}\ \emph {et~al.}(2015)\citenamefont {Keimer},
  \citenamefont {Kivelson}, \citenamefont {Norman}, \citenamefont {Uchida},\
  and\ \citenamefont {Zaanen}}]{Keimer_Nat_2015}%
  \BibitemOpen
  \bibfield  {author} {\bibinfo {author} {\bibfnamefont {B.}~\bibnamefont
  {Keimer}}, \bibinfo {author} {\bibfnamefont {S.~A.}\ \bibnamefont
  {Kivelson}}, \bibinfo {author} {\bibfnamefont {M.~R.}\ \bibnamefont
  {Norman}}, \bibinfo {author} {\bibfnamefont {S.}~\bibnamefont {Uchida}}, \
  and\ \bibinfo {author} {\bibfnamefont {J.}~\bibnamefont {Zaanen}},\ }\href
  {\doibase 10.1038/nature14165} {\bibfield  {journal} {\bibinfo  {journal}
  {Nature}\ }\textbf {\bibinfo {volume} {518}},\ \bibinfo {pages} {179}
  (\bibinfo {year} {2015})}\BibitemShut {NoStop}%
\bibitem [{\citenamefont {Emery}\ and\ \citenamefont
  {Kivelson}(1995{\natexlab{a}})}]{Emery_Nat_1995}%
  \BibitemOpen
  \bibfield  {author} {\bibinfo {author} {\bibfnamefont {V.~J.}\ \bibnamefont
  {Emery}}\ and\ \bibinfo {author} {\bibfnamefont {S.~A.}\ \bibnamefont
  {Kivelson}},\ }\href {\doibase 10.1038/374434a0} {\bibfield  {journal}
  {\bibinfo  {journal} {Nature}\ }\textbf {\bibinfo {volume} {374}},\ \bibinfo
  {pages} {434} (\bibinfo {year} {1995}{\natexlab{a}})}\BibitemShut {NoStop}%
\bibitem [{\citenamefont {Plat\'e}\ \emph {et~al.}(2005)\citenamefont
  {Plat\'e}, \citenamefont {Mottershead}, \citenamefont {Elfimov},
  \citenamefont {Peets}, \citenamefont {Liang}, \citenamefont {Bonn},
  \citenamefont {Hardy}, \citenamefont {Chiuzbaian}, \citenamefont {Falub},
  \citenamefont {Shi}, \citenamefont {Patthey},\ and\ \citenamefont
  {Damascelli}}]{Plate_PRL_2005}%
  \BibitemOpen
  \bibfield  {author} {\bibinfo {author} {\bibfnamefont {M.}~\bibnamefont
  {Plat\'e}}, \bibinfo {author} {\bibfnamefont {J.~D.~F.}\ \bibnamefont
  {Mottershead}}, \bibinfo {author} {\bibfnamefont {I.~S.}\ \bibnamefont
  {Elfimov}}, \bibinfo {author} {\bibfnamefont {D.~C.}\ \bibnamefont {Peets}},
  \bibinfo {author} {\bibfnamefont {R.}~\bibnamefont {Liang}}, \bibinfo
  {author} {\bibfnamefont {D.~A.}\ \bibnamefont {Bonn}}, \bibinfo {author}
  {\bibfnamefont {W.~N.}\ \bibnamefont {Hardy}}, \bibinfo {author}
  {\bibfnamefont {S.}~\bibnamefont {Chiuzbaian}}, \bibinfo {author}
  {\bibfnamefont {M.}~\bibnamefont {Falub}}, \bibinfo {author} {\bibfnamefont
  {M.}~\bibnamefont {Shi}}, \bibinfo {author} {\bibfnamefont {L.}~\bibnamefont
  {Patthey}}, \ and\ \bibinfo {author} {\bibfnamefont {A.}~\bibnamefont
  {Damascelli}},\ }\href {\doibase 10.1103/PhysRevLett.95.077001} {\bibfield
  {journal} {\bibinfo  {journal} {Phys. Rev. Lett.}\ }\textbf {\bibinfo
  {volume} {95}},\ \bibinfo {pages} {077001} (\bibinfo {year}
  {2005})}\BibitemShut {NoStop}%
\bibitem [{\citenamefont {Vignolle}\ \emph {et~al.}(2008)\citenamefont
  {Vignolle}, \citenamefont {Carrington}, \citenamefont {Cooper}, \citenamefont
  {French}, \citenamefont {Mackenzie}, \citenamefont {Jaudet}, \citenamefont
  {Vignolles}, \citenamefont {Proust},\ and\ \citenamefont
  {Hussey}}]{Vignolle_Nat_2008}%
  \BibitemOpen
  \bibfield  {author} {\bibinfo {author} {\bibfnamefont {B.}~\bibnamefont
  {Vignolle}}, \bibinfo {author} {\bibfnamefont {A.}~\bibnamefont
  {Carrington}}, \bibinfo {author} {\bibfnamefont {R.~A.}\ \bibnamefont
  {Cooper}}, \bibinfo {author} {\bibfnamefont {M.~M.~J.}\ \bibnamefont
  {French}}, \bibinfo {author} {\bibfnamefont {A.~P.}\ \bibnamefont
  {Mackenzie}}, \bibinfo {author} {\bibfnamefont {C.}~\bibnamefont {Jaudet}},
  \bibinfo {author} {\bibfnamefont {D.}~\bibnamefont {Vignolles}}, \bibinfo
  {author} {\bibfnamefont {C.}~\bibnamefont {Proust}}, \ and\ \bibinfo {author}
  {\bibfnamefont {N.~E.}\ \bibnamefont {Hussey}},\ }\href {\doibase
  10.1038/nature07323} {\bibfield  {journal} {\bibinfo  {journal} {Nature}\
  }\textbf {\bibinfo {volume} {455}},\ \bibinfo {pages} {952} (\bibinfo {year}
  {2008})}\BibitemShut {NoStop}%
\bibitem [{\citenamefont {Uemura}\ \emph {et~al.}(1993)\citenamefont {Uemura},
  \citenamefont {Keren}, \citenamefont {Le}, \citenamefont {Luke},
  \citenamefont {Wu}, \citenamefont {Kubo}, \citenamefont {Manako},
  \citenamefont {Shimakawa}, \citenamefont {Subramanian}, \citenamefont
  {Cobb},\ and\ \citenamefont {Markert}}]{Uemura_Nat_1993}%
  \BibitemOpen
  \bibfield  {author} {\bibinfo {author} {\bibfnamefont {Y.~J.}\ \bibnamefont
  {Uemura}}, \bibinfo {author} {\bibfnamefont {A.}~\bibnamefont {Keren}},
  \bibinfo {author} {\bibfnamefont {L.~P.}\ \bibnamefont {Le}}, \bibinfo
  {author} {\bibfnamefont {G.~M.}\ \bibnamefont {Luke}}, \bibinfo {author}
  {\bibfnamefont {W.~D.}\ \bibnamefont {Wu}}, \bibinfo {author} {\bibfnamefont
  {Y.}~\bibnamefont {Kubo}}, \bibinfo {author} {\bibfnamefont {T.}~\bibnamefont
  {Manako}}, \bibinfo {author} {\bibfnamefont {Y.}~\bibnamefont {Shimakawa}},
  \bibinfo {author} {\bibfnamefont {M.}~\bibnamefont {Subramanian}}, \bibinfo
  {author} {\bibfnamefont {J.~L.}\ \bibnamefont {Cobb}}, \ and\ \bibinfo
  {author} {\bibfnamefont {J.~T.}\ \bibnamefont {Markert}},\ }\href {\doibase
  10.1038/364605a0} {\bibfield  {journal} {\bibinfo  {journal} {Nature}\
  }\textbf {\bibinfo {volume} {364}},\ \bibinfo {pages} {605} (\bibinfo {year}
  {1993})}\BibitemShut {NoStop}%
\bibitem [{\citenamefont {Niedermayer}\ \emph {et~al.}(1993)\citenamefont
  {Niedermayer}, \citenamefont {Bernhard}, \citenamefont {Binninger},
  \citenamefont {Gl\"uckler}, \citenamefont {Tallon}, \citenamefont {Ansaldo},\
  and\ \citenamefont {Budnick}}]{Niedermayer_PRL_1993}%
  \BibitemOpen
  \bibfield  {author} {\bibinfo {author} {\bibfnamefont {C.}~\bibnamefont
  {Niedermayer}}, \bibinfo {author} {\bibfnamefont {C.}~\bibnamefont
  {Bernhard}}, \bibinfo {author} {\bibfnamefont {U.}~\bibnamefont {Binninger}},
  \bibinfo {author} {\bibfnamefont {H.}~\bibnamefont {Gl\"uckler}}, \bibinfo
  {author} {\bibfnamefont {J.~L.}\ \bibnamefont {Tallon}}, \bibinfo {author}
  {\bibfnamefont {E.~J.}\ \bibnamefont {Ansaldo}}, \ and\ \bibinfo {author}
  {\bibfnamefont {J.~I.}\ \bibnamefont {Budnick}},\ }\href {\doibase
  10.1103/PhysRevLett.71.1764} {\bibfield  {journal} {\bibinfo  {journal}
  {Phys. Rev. Lett.}\ }\textbf {\bibinfo {volume} {71}},\ \bibinfo {pages}
  {1764} (\bibinfo {year} {1993})}\BibitemShut {NoStop}%
\bibitem [{\citenamefont {Panagopoulos}\ \emph {et~al.}(2003)\citenamefont
  {Panagopoulos}, \citenamefont {Xiang}, \citenamefont {Anukool}, \citenamefont
  {Cooper}, \citenamefont {Wang},\ and\ \citenamefont
  {Chu}}]{Panagopoulos_PRB_2003}%
  \BibitemOpen
  \bibfield  {author} {\bibinfo {author} {\bibfnamefont {C.}~\bibnamefont
  {Panagopoulos}}, \bibinfo {author} {\bibfnamefont {T.}~\bibnamefont {Xiang}},
  \bibinfo {author} {\bibfnamefont {W.}~\bibnamefont {Anukool}}, \bibinfo
  {author} {\bibfnamefont {J.~R.}\ \bibnamefont {Cooper}}, \bibinfo {author}
  {\bibfnamefont {Y.~S.}\ \bibnamefont {Wang}}, \ and\ \bibinfo {author}
  {\bibfnamefont {C.~W.}\ \bibnamefont {Chu}},\ }\href {\doibase
  10.1103/PhysRevB.67.220502} {\bibfield  {journal} {\bibinfo  {journal} {Phys.
  Rev. B}\ }\textbf {\bibinfo {volume} {67}},\ \bibinfo {pages} {220502}
  (\bibinfo {year} {2003})}\BibitemShut {NoStop}%
\bibitem [{\citenamefont {Corson}\ \emph
  {et~al.}(2000{\natexlab{a}})\citenamefont {Corson}, \citenamefont
  {Orenstein}, \citenamefont {Eckstein},\ and\ \citenamefont
  {Bo{\v{z}}ovi{\'{c}}}}]{Corson_PhysicaB_2000}%
  \BibitemOpen
  \bibfield  {author} {\bibinfo {author} {\bibfnamefont {J.}~\bibnamefont
  {Corson}}, \bibinfo {author} {\bibfnamefont {J.}~\bibnamefont {Orenstein}},
  \bibinfo {author} {\bibfnamefont {J.}~\bibnamefont {Eckstein}}, \ and\
  \bibinfo {author} {\bibfnamefont {I.}~\bibnamefont {Bo{\v{z}}ovi{\'{c}}}},\
  }\href {\doibase https://doi.org/10.1016/S0921-4526(99)01580-X} {\bibfield
  {journal} {\bibinfo  {journal} {Physica B: Condensed Matter}\ }\textbf
  {\bibinfo {volume} {280}},\ \bibinfo {pages} {212 } (\bibinfo {year}
  {2000}{\natexlab{a}})}\BibitemShut {NoStop}%
\bibitem [{\citenamefont {Locquet}\ \emph {et~al.}(1996)\citenamefont
  {Locquet}, \citenamefont {Jaccard}, \citenamefont {Cretton}, \citenamefont
  {Williams}, \citenamefont {Arrouy}, \citenamefont {M\"achler}, \citenamefont
  {Schneider}, \citenamefont {Fischer},\ and\ \citenamefont
  {Martinoli}}]{Locquet_PRB_54}%
  \BibitemOpen
  \bibfield  {author} {\bibinfo {author} {\bibfnamefont {J.-P.}\ \bibnamefont
  {Locquet}}, \bibinfo {author} {\bibfnamefont {Y.}~\bibnamefont {Jaccard}},
  \bibinfo {author} {\bibfnamefont {A.}~\bibnamefont {Cretton}}, \bibinfo
  {author} {\bibfnamefont {E.~J.}\ \bibnamefont {Williams}}, \bibinfo {author}
  {\bibfnamefont {F.}~\bibnamefont {Arrouy}}, \bibinfo {author} {\bibfnamefont
  {E.}~\bibnamefont {M\"achler}}, \bibinfo {author} {\bibfnamefont
  {T.}~\bibnamefont {Schneider}}, \bibinfo {author} {\bibfnamefont
  {O.}~\bibnamefont {Fischer}}, \ and\ \bibinfo {author} {\bibfnamefont
  {P.}~\bibnamefont {Martinoli}},\ }\href {\doibase 10.1103/PhysRevB.54.7481}
  {\bibfield  {journal} {\bibinfo  {journal} {Phys. Rev. B}\ }\textbf {\bibinfo
  {volume} {54}},\ \bibinfo {pages} {7481} (\bibinfo {year}
  {1996})}\BibitemShut {NoStop}%
\bibitem [{\citenamefont {Lemberger}\ \emph {et~al.}(2010)\citenamefont
  {Lemberger}, \citenamefont {Hetel}, \citenamefont {Tsukada},\ and\
  \citenamefont {Naito}}]{Lemberger_PRB_2010}%
  \BibitemOpen
  \bibfield  {author} {\bibinfo {author} {\bibfnamefont {T.~R.}\ \bibnamefont
  {Lemberger}}, \bibinfo {author} {\bibfnamefont {I.}~\bibnamefont {Hetel}},
  \bibinfo {author} {\bibfnamefont {A.}~\bibnamefont {Tsukada}}, \ and\
  \bibinfo {author} {\bibfnamefont {M.}~\bibnamefont {Naito}},\ }\href
  {\doibase 10.1103/PhysRevB.82.214513} {\bibfield  {journal} {\bibinfo
  {journal} {Phys. Rev. B}\ }\textbf {\bibinfo {volume} {82}},\ \bibinfo
  {pages} {214513} (\bibinfo {year} {2010})}\BibitemShut {NoStop}%
\bibitem [{\citenamefont {Lemberger}\ \emph {et~al.}(2011)\citenamefont
  {Lemberger}, \citenamefont {Hetel}, \citenamefont {Tsukada}, \citenamefont
  {Naito},\ and\ \citenamefont {Randeria}}]{Lemberger_PRB_2011}%
  \BibitemOpen
  \bibfield  {author} {\bibinfo {author} {\bibfnamefont {T.~R.}\ \bibnamefont
  {Lemberger}}, \bibinfo {author} {\bibfnamefont {I.}~\bibnamefont {Hetel}},
  \bibinfo {author} {\bibfnamefont {A.}~\bibnamefont {Tsukada}}, \bibinfo
  {author} {\bibfnamefont {M.}~\bibnamefont {Naito}}, \ and\ \bibinfo {author}
  {\bibfnamefont {M.}~\bibnamefont {Randeria}},\ }\href {\doibase
  10.1103/PhysRevB.83.140507} {\bibfield  {journal} {\bibinfo  {journal} {Phys.
  Rev. B}\ }\textbf {\bibinfo {volume} {83}},\ \bibinfo {pages} {140507}
  (\bibinfo {year} {2011})}\BibitemShut {NoStop}%
\bibitem [{\citenamefont {Rourke}\ \emph {et~al.}(2011)\citenamefont {Rourke},
  \citenamefont {Mouzopoulou}, \citenamefont {Xu}, \citenamefont
  {Panagopoulos}, \citenamefont {Wang}, \citenamefont {Vignolle}, \citenamefont
  {Proust}, \citenamefont {Kurganova}, \citenamefont {Zeitler}, \citenamefont
  {Tanabe}, \citenamefont {Adachi}, \citenamefont {Koike},\ and\ \citenamefont
  {Hussey}}]{Rourke_Natphys_2011}%
  \BibitemOpen
  \bibfield  {author} {\bibinfo {author} {\bibfnamefont {P.~M.~C.}\
  \bibnamefont {Rourke}}, \bibinfo {author} {\bibfnamefont {I.}~\bibnamefont
  {Mouzopoulou}}, \bibinfo {author} {\bibfnamefont {X.}~\bibnamefont {Xu}},
  \bibinfo {author} {\bibfnamefont {C.}~\bibnamefont {Panagopoulos}}, \bibinfo
  {author} {\bibfnamefont {Y.}~\bibnamefont {Wang}}, \bibinfo {author}
  {\bibfnamefont {B.}~\bibnamefont {Vignolle}}, \bibinfo {author}
  {\bibfnamefont {C.}~\bibnamefont {Proust}}, \bibinfo {author} {\bibfnamefont
  {E.~V.}\ \bibnamefont {Kurganova}}, \bibinfo {author} {\bibfnamefont
  {U.}~\bibnamefont {Zeitler}}, \bibinfo {author} {\bibfnamefont
  {Y.}~\bibnamefont {Tanabe}}, \bibinfo {author} {\bibfnamefont
  {T.}~\bibnamefont {Adachi}}, \bibinfo {author} {\bibfnamefont
  {Y.}~\bibnamefont {Koike}}, \ and\ \bibinfo {author} {\bibfnamefont {N.~E.}\
  \bibnamefont {Hussey}},\ }\href {\doibase 10.1038/nphys1945} {\bibfield
  {journal} {\bibinfo  {journal} {Nature Physics}\ }\textbf {\bibinfo {volume}
  {7}},\ \bibinfo {pages} {455} (\bibinfo {year} {2011})}\BibitemShut {NoStop}%
\bibitem [{\citenamefont {Bo{\v{z}}ovi{\'{c}}}\ \emph
  {et~al.}(2016)\citenamefont {Bo{\v{z}}ovi{\'{c}}}, \citenamefont {He},
  \citenamefont {Wu},\ and\ \citenamefont {Bollinger}}]{Bozovic_Nat_2016}%
  \BibitemOpen
  \bibfield  {author} {\bibinfo {author} {\bibfnamefont {I.}~\bibnamefont
  {Bo{\v{z}}ovi{\'{c}}}}, \bibinfo {author} {\bibfnamefont {X.}~\bibnamefont
  {He}}, \bibinfo {author} {\bibfnamefont {J.}~\bibnamefont {Wu}}, \ and\
  \bibinfo {author} {\bibfnamefont {A.~T.}\ \bibnamefont {Bollinger}},\ }\href
  {\doibase 10.1038/nature19061} {\bibfield  {journal} {\bibinfo  {journal}
  {Nature}\ }\textbf {\bibinfo {volume} {536}},\ \bibinfo {pages} {309}
  (\bibinfo {year} {2016})}\BibitemShut {NoStop}%
\bibitem [{\citenamefont {Corson}\ \emph {et~al.}(1999)\citenamefont {Corson},
  \citenamefont {Mallozzi}, \citenamefont {Orenstein}, \citenamefont
  {Eckstein},\ and\ \citenamefont {Bo{\v{z}}ovi{\'{c}}}}]{Corson_Nature_1999}%
  \BibitemOpen
  \bibfield  {author} {\bibinfo {author} {\bibfnamefont {J.}~\bibnamefont
  {Corson}}, \bibinfo {author} {\bibfnamefont {R.}~\bibnamefont {Mallozzi}},
  \bibinfo {author} {\bibfnamefont {J.}~\bibnamefont {Orenstein}}, \bibinfo
  {author} {\bibfnamefont {J.~N.}\ \bibnamefont {Eckstein}}, \ and\ \bibinfo
  {author} {\bibfnamefont {I.}~\bibnamefont {Bo{\v{z}}ovi{\'{c}}}},\ }\href
  {\doibase 10.1038/18402} {\bibfield  {journal} {\bibinfo  {journal} {Nature}\
  }\textbf {\bibinfo {volume} {398}},\ \bibinfo {pages} {221} (\bibinfo {year}
  {1999})}\BibitemShut {NoStop}%
\bibitem [{\citenamefont {Corson}\ \emph
  {et~al.}(2000{\natexlab{b}})\citenamefont {Corson}, \citenamefont
  {Orenstein}, \citenamefont {Oh}, \citenamefont {O'Donnell},\ and\
  \citenamefont {Eckstein}}]{Corson_PRL_2000}%
  \BibitemOpen
  \bibfield  {author} {\bibinfo {author} {\bibfnamefont {J.}~\bibnamefont
  {Corson}}, \bibinfo {author} {\bibfnamefont {J.}~\bibnamefont {Orenstein}},
  \bibinfo {author} {\bibfnamefont {S.}~\bibnamefont {Oh}}, \bibinfo {author}
  {\bibfnamefont {J.}~\bibnamefont {O'Donnell}}, \ and\ \bibinfo {author}
  {\bibfnamefont {J.~N.}\ \bibnamefont {Eckstein}},\ }\href {\doibase
  10.1103/PhysRevLett.85.2569} {\bibfield  {journal} {\bibinfo  {journal}
  {Phys. Rev. Lett.}\ }\textbf {\bibinfo {volume} {85}},\ \bibinfo {pages}
  {2569} (\bibinfo {year} {2000}{\natexlab{b}})}\BibitemShut {NoStop}%
\bibitem [{\citenamefont {Molegraaf}\ \emph {et~al.}(2002)\citenamefont
  {Molegraaf}, \citenamefont {Presura}, \citenamefont {van~der Marel},
  \citenamefont {Kes},\ and\ \citenamefont
  {Li}}]{molegraaf2002superconductivity}%
  \BibitemOpen
  \bibfield  {author} {\bibinfo {author} {\bibfnamefont {H.}~\bibnamefont
  {Molegraaf}}, \bibinfo {author} {\bibfnamefont {C.}~\bibnamefont {Presura}},
  \bibinfo {author} {\bibfnamefont {D.}~\bibnamefont {van~der Marel}}, \bibinfo
  {author} {\bibfnamefont {P.}~\bibnamefont {Kes}}, \ and\ \bibinfo {author}
  {\bibfnamefont {M.}~\bibnamefont {Li}},\ }\href@noop {} {\bibfield  {journal}
  {\bibinfo  {journal} {Science}\ }\textbf {\bibinfo {volume} {295}},\ \bibinfo
  {pages} {2239} (\bibinfo {year} {2002})}\BibitemShut {NoStop}%
\bibitem [{\citenamefont {Hirschfeld}\ \emph {et~al.}(1993)\citenamefont
  {Hirschfeld}, \citenamefont {Putikka},\ and\ \citenamefont
  {Scalapino}}]{Hirschfeld_PRL_1993}%
  \BibitemOpen
  \bibfield  {author} {\bibinfo {author} {\bibfnamefont {P.~J.}\ \bibnamefont
  {Hirschfeld}}, \bibinfo {author} {\bibfnamefont {W.~O.}\ \bibnamefont
  {Putikka}}, \ and\ \bibinfo {author} {\bibfnamefont {D.~J.}\ \bibnamefont
  {Scalapino}},\ }\href {\doibase 10.1103/PhysRevLett.71.3705} {\bibfield
  {journal} {\bibinfo  {journal} {Phys. Rev. Lett.}\ }\textbf {\bibinfo
  {volume} {71}},\ \bibinfo {pages} {3705} (\bibinfo {year}
  {1993})}\BibitemShut {NoStop}%
\bibitem [{\citenamefont {Hirschfeld}\ and\ \citenamefont
  {Goldenfeld}(1993)}]{Hirschfeld_PRB_1993}%
  \BibitemOpen
  \bibfield  {author} {\bibinfo {author} {\bibfnamefont {P.~J.}\ \bibnamefont
  {Hirschfeld}}\ and\ \bibinfo {author} {\bibfnamefont {N.}~\bibnamefont
  {Goldenfeld}},\ }\href {\doibase 10.1103/PhysRevB.48.4219} {\bibfield
  {journal} {\bibinfo  {journal} {Phys. Rev. B}\ }\textbf {\bibinfo {volume}
  {48}},\ \bibinfo {pages} {4219} (\bibinfo {year} {1993})}\BibitemShut
  {NoStop}%
\bibitem [{\citenamefont {Hosseini}\ \emph {et~al.}(1999)\citenamefont
  {Hosseini}, \citenamefont {Harris}, \citenamefont {Kamal}, \citenamefont
  {Dosanjh}, \citenamefont {Preston}, \citenamefont {Liang}, \citenamefont
  {Hardy},\ and\ \citenamefont {Bonn}}]{Hosseini_PRB_1999}%
  \BibitemOpen
  \bibfield  {author} {\bibinfo {author} {\bibfnamefont {A.}~\bibnamefont
  {Hosseini}}, \bibinfo {author} {\bibfnamefont {R.}~\bibnamefont {Harris}},
  \bibinfo {author} {\bibfnamefont {S.}~\bibnamefont {Kamal}}, \bibinfo
  {author} {\bibfnamefont {P.}~\bibnamefont {Dosanjh}}, \bibinfo {author}
  {\bibfnamefont {J.}~\bibnamefont {Preston}}, \bibinfo {author} {\bibfnamefont
  {R.}~\bibnamefont {Liang}}, \bibinfo {author} {\bibfnamefont {W.~N.}\
  \bibnamefont {Hardy}}, \ and\ \bibinfo {author} {\bibfnamefont {D.~A.}\
  \bibnamefont {Bonn}},\ }\href {\doibase 10.1103/PhysRevB.60.1349} {\bibfield
  {journal} {\bibinfo  {journal} {Phys. Rev. B}\ }\textbf {\bibinfo {volume}
  {60}},\ \bibinfo {pages} {1349} (\bibinfo {year} {1999})}\BibitemShut
  {NoStop}%
\bibitem [{\citenamefont {Durst}\ and\ \citenamefont
  {Lee}(2000)}]{Durst_PRB_2000}%
  \BibitemOpen
  \bibfield  {author} {\bibinfo {author} {\bibfnamefont {A.~C.}\ \bibnamefont
  {Durst}}\ and\ \bibinfo {author} {\bibfnamefont {P.~A.}\ \bibnamefont
  {Lee}},\ }\href {\doibase 10.1103/PhysRevB.62.1270} {\bibfield  {journal}
  {\bibinfo  {journal} {Phys. Rev. B}\ }\textbf {\bibinfo {volume} {62}},\
  \bibinfo {pages} {1270} (\bibinfo {year} {2000})}\BibitemShut {NoStop}%
\bibitem [{\citenamefont {Broun}\ \emph {et~al.}(2007)\citenamefont {Broun},
  \citenamefont {Huttema}, \citenamefont {Turner}, \citenamefont {\"Ozcan},
  \citenamefont {Morgan}, \citenamefont {Liang}, \citenamefont {Hardy},\ and\
  \citenamefont {Bonn}}]{Broun_PRL_2007}%
  \BibitemOpen
  \bibfield  {author} {\bibinfo {author} {\bibfnamefont {D.~M.}\ \bibnamefont
  {Broun}}, \bibinfo {author} {\bibfnamefont {W.~A.}\ \bibnamefont {Huttema}},
  \bibinfo {author} {\bibfnamefont {P.~J.}\ \bibnamefont {Turner}}, \bibinfo
  {author} {\bibfnamefont {S.}~\bibnamefont {\"Ozcan}}, \bibinfo {author}
  {\bibfnamefont {B.}~\bibnamefont {Morgan}}, \bibinfo {author} {\bibfnamefont
  {R.}~\bibnamefont {Liang}}, \bibinfo {author} {\bibfnamefont {W.~N.}\
  \bibnamefont {Hardy}}, \ and\ \bibinfo {author} {\bibfnamefont {D.~A.}\
  \bibnamefont {Bonn}},\ }\href {\doibase 10.1103/PhysRevLett.99.237003}
  {\bibfield  {journal} {\bibinfo  {journal} {Phys. Rev. Lett.}\ }\textbf
  {\bibinfo {volume} {99}},\ \bibinfo {pages} {237003} (\bibinfo {year}
  {2007})}\BibitemShut {NoStop}%
\bibitem [{\citenamefont {Lee-Hone}\ \emph {et~al.}(2017)\citenamefont
  {Lee-Hone}, \citenamefont {Dodge},\ and\ \citenamefont
  {Broun}}]{Lee-Hone_PRB_2017}%
  \BibitemOpen
  \bibfield  {author} {\bibinfo {author} {\bibfnamefont {N.~R.}\ \bibnamefont
  {Lee-Hone}}, \bibinfo {author} {\bibfnamefont {J.~S.}\ \bibnamefont {Dodge}},
  \ and\ \bibinfo {author} {\bibfnamefont {D.~M.}\ \bibnamefont {Broun}},\
  }\href {\doibase 10.1103/PhysRevB.96.024501} {\bibfield  {journal} {\bibinfo
  {journal} {Phys. Rev. B}\ }\textbf {\bibinfo {volume} {96}},\ \bibinfo
  {pages} {024501} (\bibinfo {year} {2017})}\BibitemShut {NoStop}%
\bibitem [{\citenamefont {Won}\ and\ \citenamefont
  {Maki}(1994)}]{Maki_PRB_1994}%
  \BibitemOpen
  \bibfield  {author} {\bibinfo {author} {\bibfnamefont {H.}~\bibnamefont
  {Won}}\ and\ \bibinfo {author} {\bibfnamefont {K.}~\bibnamefont {Maki}},\
  }\href {\doibase 10.1103/PhysRevB.49.1397} {\bibfield  {journal} {\bibinfo
  {journal} {Phys. Rev. B}\ }\textbf {\bibinfo {volume} {49}},\ \bibinfo
  {pages} {1397} (\bibinfo {year} {1994})}\BibitemShut {NoStop}%
\bibitem [{\citenamefont {Tajima}\ \emph {et~al.}(2005)\citenamefont {Tajima},
  \citenamefont {Fudamoto}, \citenamefont {Kakeshita}, \citenamefont
  {Gorshunov}, \citenamefont {\ifmmode~\check{Z}\else \v{Z}\fi{}elezn\'y},
  \citenamefont {Kojima}, \citenamefont {Dressel},\ and\ \citenamefont
  {Uchida}}]{Tajima_PRB_2005}%
  \BibitemOpen
  \bibfield  {author} {\bibinfo {author} {\bibfnamefont {S.}~\bibnamefont
  {Tajima}}, \bibinfo {author} {\bibfnamefont {Y.}~\bibnamefont {Fudamoto}},
  \bibinfo {author} {\bibfnamefont {T.}~\bibnamefont {Kakeshita}}, \bibinfo
  {author} {\bibfnamefont {B.}~\bibnamefont {Gorshunov}}, \bibinfo {author}
  {\bibfnamefont {V.}~\bibnamefont {\ifmmode~\check{Z}\else
  \v{Z}\fi{}elezn\'y}}, \bibinfo {author} {\bibfnamefont {K.~M.}\ \bibnamefont
  {Kojima}}, \bibinfo {author} {\bibfnamefont {M.}~\bibnamefont {Dressel}}, \
  and\ \bibinfo {author} {\bibfnamefont {S.}~\bibnamefont {Uchida}},\ }\href
  {\doibase 10.1103/PhysRevB.71.094508} {\bibfield  {journal} {\bibinfo
  {journal} {Phys. Rev. B}\ }\textbf {\bibinfo {volume} {71}},\ \bibinfo
  {pages} {094508} (\bibinfo {year} {2005})}\BibitemShut {NoStop}%
\bibitem [{\citenamefont {Lee-Hone}\ \emph {et~al.}(2018)\citenamefont
  {Lee-Hone}, \citenamefont {Mishra}, \citenamefont {Broun},\ and\
  \citenamefont {Hirschfeld}}]{Lee-Hone_arxiv_2018}%
  \BibitemOpen
  \bibfield  {author} {\bibinfo {author} {\bibfnamefont {N.~R.}\ \bibnamefont
  {Lee-Hone}}, \bibinfo {author} {\bibfnamefont {V.}~\bibnamefont {Mishra}},
  \bibinfo {author} {\bibfnamefont {D.~M.}\ \bibnamefont {Broun}}, \ and\
  \bibinfo {author} {\bibfnamefont {P.~J.}\ \bibnamefont {Hirschfeld}},\
  }\href@noop {} {\bibfield  {journal} {\bibinfo  {journal} {arXiv:1802.10198}\
  } (\bibinfo {year} {2018})}\BibitemShut {NoStop}%
\bibitem [{\citenamefont {Wang}\ \emph {et~al.}(2007)\citenamefont {Wang},
  \citenamefont {Yan}, \citenamefont {Shan}, \citenamefont {Wen}, \citenamefont
  {Tanabe}, \citenamefont {Adachi},\ and\ \citenamefont {Koike}}]{Wang_PRB_76}%
  \BibitemOpen
  \bibfield  {author} {\bibinfo {author} {\bibfnamefont {Y.}~\bibnamefont
  {Wang}}, \bibinfo {author} {\bibfnamefont {J.}~\bibnamefont {Yan}}, \bibinfo
  {author} {\bibfnamefont {L.}~\bibnamefont {Shan}}, \bibinfo {author}
  {\bibfnamefont {H.-H.}\ \bibnamefont {Wen}}, \bibinfo {author} {\bibfnamefont
  {Y.}~\bibnamefont {Tanabe}}, \bibinfo {author} {\bibfnamefont
  {T.}~\bibnamefont {Adachi}}, \ and\ \bibinfo {author} {\bibfnamefont
  {Y.}~\bibnamefont {Koike}},\ }\href {\doibase 10.1103/PhysRevB.76.064512}
  {\bibfield  {journal} {\bibinfo  {journal} {Phys. Rev. B}\ }\textbf {\bibinfo
  {volume} {76}},\ \bibinfo {pages} {064512} (\bibinfo {year}
  {2007})}\BibitemShut {NoStop}%
\bibitem [{\citenamefont {Bilbro}\ \emph
  {et~al.}(2011{\natexlab{a}})\citenamefont {Bilbro}, \citenamefont {Aguilar},
  \citenamefont {Logvenov}, \citenamefont {Pelleg}, \citenamefont
  {Bo{\v{z}}ovi{\'{c}}},\ and\ \citenamefont {Armitage}}]{Bilbro_Natphys_2011}%
  \BibitemOpen
  \bibfield  {author} {\bibinfo {author} {\bibfnamefont {L.~S.}\ \bibnamefont
  {Bilbro}}, \bibinfo {author} {\bibfnamefont {R.~V.}\ \bibnamefont {Aguilar}},
  \bibinfo {author} {\bibfnamefont {G.}~\bibnamefont {Logvenov}}, \bibinfo
  {author} {\bibfnamefont {O.}~\bibnamefont {Pelleg}}, \bibinfo {author}
  {\bibfnamefont {I.}~\bibnamefont {Bo{\v{z}}ovi{\'{c}}}}, \ and\ \bibinfo
  {author} {\bibfnamefont {N.~P.}\ \bibnamefont {Armitage}},\ }\href {\doibase
  10.1038/nphys1912} {\bibfield  {journal} {\bibinfo  {journal} {Nature
  Physics}\ }\textbf {\bibinfo {volume} {7}},\ \bibinfo {pages} {298} (\bibinfo
  {year} {2011}{\natexlab{a}})}\BibitemShut {NoStop}%
\bibitem [{\citenamefont {Bilbro}\ \emph
  {et~al.}(2011{\natexlab{b}})\citenamefont {Bilbro}, \citenamefont
  {Vald\'es~Aguilar}, \citenamefont {Logvenov}, \citenamefont {Bozovic},\ and\
  \citenamefont {Armitage}}]{Bilbro11a}%
  \BibitemOpen
  \bibfield  {author} {\bibinfo {author} {\bibfnamefont {L.~S.}\ \bibnamefont
  {Bilbro}}, \bibinfo {author} {\bibfnamefont {R.}~\bibnamefont
  {Vald\'es~Aguilar}}, \bibinfo {author} {\bibfnamefont {G.}~\bibnamefont
  {Logvenov}}, \bibinfo {author} {\bibfnamefont {I.}~\bibnamefont {Bozovic}}, \
  and\ \bibinfo {author} {\bibfnamefont {N.~P.}\ \bibnamefont {Armitage}},\
  }\href {\doibase 10.1103/PhysRevB.84.100511} {\bibfield  {journal} {\bibinfo
  {journal} {Phys. Rev. B}\ }\textbf {\bibinfo {volume} {84}},\ \bibinfo
  {pages} {100511} (\bibinfo {year} {2011}{\natexlab{b}})}\BibitemShut
  {NoStop}%
\bibitem [{\citenamefont {Ioffe}(1999)}]{Ioffe_Science_1999}%
  \BibitemOpen
  \bibfield  {author} {\bibinfo {author} {\bibfnamefont {L.~B.}\ \bibnamefont
  {Ioffe}},\ }\href {\doibase 10.1126/science.285.5431.1241} {\bibfield
  {journal} {\bibinfo  {journal} {Science}\ }\textbf {\bibinfo {volume}
  {285}},\ \bibinfo {pages} {1241} (\bibinfo {year} {1999})}\BibitemShut
  {NoStop}%
\bibitem [{\citenamefont {Benfatto}\ \emph {et~al.}(2001)\citenamefont
  {Benfatto}, \citenamefont {Caprara}, \citenamefont {Castellani},
  \citenamefont {Paramekanti},\ and\ \citenamefont
  {Randeria}}]{Benfatto_PRB_2001}%
  \BibitemOpen
  \bibfield  {author} {\bibinfo {author} {\bibfnamefont {L.}~\bibnamefont
  {Benfatto}}, \bibinfo {author} {\bibfnamefont {S.}~\bibnamefont {Caprara}},
  \bibinfo {author} {\bibfnamefont {C.}~\bibnamefont {Castellani}}, \bibinfo
  {author} {\bibfnamefont {A.}~\bibnamefont {Paramekanti}}, \ and\ \bibinfo
  {author} {\bibfnamefont {M.}~\bibnamefont {Randeria}},\ }\href {\doibase
  10.1103/PhysRevB.63.174513} {\bibfield  {journal} {\bibinfo  {journal} {Phys.
  Rev. B}\ }\textbf {\bibinfo {volume} {63}},\ \bibinfo {pages} {174513}
  (\bibinfo {year} {2001})}\BibitemShut {NoStop}%
\bibitem [{\citenamefont {Kwon}\ \emph {et~al.}(2001)\citenamefont {Kwon},
  \citenamefont {Dorsey},\ and\ \citenamefont {Hirschfeld}}]{Kwon_PRL_2001}%
  \BibitemOpen
  \bibfield  {author} {\bibinfo {author} {\bibfnamefont {H.-J.}\ \bibnamefont
  {Kwon}}, \bibinfo {author} {\bibfnamefont {A.~T.}\ \bibnamefont {Dorsey}}, \
  and\ \bibinfo {author} {\bibfnamefont {P.~J.}\ \bibnamefont {Hirschfeld}},\
  }\href {\doibase 10.1103/PhysRevLett.86.3875} {\bibfield  {journal} {\bibinfo
   {journal} {Phys. Rev. Lett.}\ }\textbf {\bibinfo {volume} {86}},\ \bibinfo
  {pages} {3875} (\bibinfo {year} {2001})}\BibitemShut {NoStop}%
\bibitem [{\citenamefont {Chakravarty}\ \emph {et~al.}(1986)\citenamefont
  {Chakravarty}, \citenamefont {Ingold}, \citenamefont {Kivelson},\ and\
  \citenamefont {Luther}}]{Chakravarty_PRL_1986}%
  \BibitemOpen
  \bibfield  {author} {\bibinfo {author} {\bibfnamefont {S.}~\bibnamefont
  {Chakravarty}}, \bibinfo {author} {\bibfnamefont {G.-L.}\ \bibnamefont
  {Ingold}}, \bibinfo {author} {\bibfnamefont {S.}~\bibnamefont {Kivelson}}, \
  and\ \bibinfo {author} {\bibfnamefont {A.}~\bibnamefont {Luther}},\ }\href
  {\doibase 10.1103/PhysRevLett.56.2303} {\bibfield  {journal} {\bibinfo
  {journal} {Phys. Rev. Lett.}\ }\textbf {\bibinfo {volume} {56}},\ \bibinfo
  {pages} {2303} (\bibinfo {year} {1986})}\BibitemShut {NoStop}%
\bibitem [{\citenamefont {Emery}\ and\ \citenamefont
  {Kivelson}(1995{\natexlab{b}})}]{Kivelson_PRL_1995}%
  \BibitemOpen
  \bibfield  {author} {\bibinfo {author} {\bibfnamefont {V.~J.}\ \bibnamefont
  {Emery}}\ and\ \bibinfo {author} {\bibfnamefont {S.~A.}\ \bibnamefont
  {Kivelson}},\ }\href {\doibase 10.1103/PhysRevLett.74.3253} {\bibfield
  {journal} {\bibinfo  {journal} {Phys. Rev. Lett.}\ }\textbf {\bibinfo
  {volume} {74}},\ \bibinfo {pages} {3253} (\bibinfo {year}
  {1995}{\natexlab{b}})}\BibitemShut {NoStop}%
\bibitem [{\citenamefont {Mondal}\ \emph {et~al.}(2011)\citenamefont {Mondal},
  \citenamefont {Kamlapure}, \citenamefont {Chand}, \citenamefont {Saraswat},
  \citenamefont {Kumar}, \citenamefont {Jesudasan}, \citenamefont {Benfatto},
  \citenamefont {Tripathi},\ and\ \citenamefont
  {Raychaudhuri}}]{Mondal_PRL_2011}%
  \BibitemOpen
  \bibfield  {author} {\bibinfo {author} {\bibfnamefont {M.}~\bibnamefont
  {Mondal}}, \bibinfo {author} {\bibfnamefont {A.}~\bibnamefont {Kamlapure}},
  \bibinfo {author} {\bibfnamefont {M.}~\bibnamefont {Chand}}, \bibinfo
  {author} {\bibfnamefont {G.}~\bibnamefont {Saraswat}}, \bibinfo {author}
  {\bibfnamefont {S.}~\bibnamefont {Kumar}}, \bibinfo {author} {\bibfnamefont
  {J.}~\bibnamefont {Jesudasan}}, \bibinfo {author} {\bibfnamefont
  {L.}~\bibnamefont {Benfatto}}, \bibinfo {author} {\bibfnamefont
  {V.}~\bibnamefont {Tripathi}}, \ and\ \bibinfo {author} {\bibfnamefont
  {P.}~\bibnamefont {Raychaudhuri}},\ }\href {\doibase
  10.1103/PhysRevLett.106.047001} {\bibfield  {journal} {\bibinfo  {journal}
  {Phys. Rev. Lett.}\ }\textbf {\bibinfo {volume} {106}},\ \bibinfo {pages}
  {047001} (\bibinfo {year} {2011})}\BibitemShut {NoStop}%
\end{thebibliography}%

\end{document}